\documentclass[a4paper,11pt]{article}
\usepackage{pos}
\usepackage[symbol]{footmisc}

\title{Searching for the Dark Photon with PADME}
\author*[a]{Kalina Dimitrova for the PADME Collaboration\footnote[2]{The PADME collaboration: 
M. Antonelli,
C. Arcangeletti,
F. Bossi, 
E. Di Meco, 
P. Gianotti, G. Mancini, M. Mancini, I. Sarra, T. Spadaro, 
E. Vilucchi, (INFN
Laboratori Nazionali di Frascati), V. Kozhuharov (Faculty of Physics, University of Sofia “St. Kl. Ohridski”, and
INFN Laboratori Nazionali di Frascati), K. Dimitrova, S. Ivanov, Sv. Ivanov, K. Kostova, R. Simeonov (Faculty of Physics,
University of Sofia “St. Kl. Ohridski”), 
F. Anulli, F. Ferrarotto, M. Garattini, E. Leonardi, P. Valente
(INFN Sezione di Roma),
M. Raggi (Physics Department, “Sapienza” Universit\`a di Roma and INFN Sezione di Roma), 
P. Massarotti (INFN Sezione di Napoli and Physics Department, “Ettore Pancini”,Universit\`a degli studi di Napoli Federico II),
A. Frankenthal (Department of Physicsand Astronomy, University of California, Irvine)}}

\affiliation[a]{Faculty of Physics, Sofia University "St. Kliment Ohridski",\\
  5 J. Bourchier Blvd., 1164 Sofia, Bulgaria}

\emailAdd{kalina@phys.uni-sofia.bg}

\abstract{The PADME Experiment at Laboratori Nationali di Frascati is designed to search for the Dark Photon, a hypothetical gauge boson responsible for the interaction between the visible and the hidden sector. PADME explores the process of annihilation of beam positrons with the electrons in a fixed target, employing the missing mass technique: in case the annihilation results in the associate production of one visible and one Dark photon, the first can be registered by the experiment's electromagnetic calorimeter and the Dark Photon mass can be reconstructed knowing the beam energy. This paper presents the analysis techniques that are being employed for the PADME data, as well as the background composition and rejection procedure.}

\FullConference{3rd General Meeting of the COST Action COSMIC WISPers (CA21106) - 3rd Training School of the COST Action COSMIC WISPers (CA21106) (COSMICWISPers2025)\\
9–12 Sept 2025 and 16–19 Sept 2025\\
Sofia, Bulgaria and Annecy, France\\}

\begin{document}
\maketitle

\section{Introduction}
%Since the early XX-th century different anomalous astrophysical and cosmological effects have been observed which has led to the introduction of the Dark Matter phenomenon. 
%Among the possible explanations is the existence of a hidden sector of particles which interact with the Standard Model particles through a new interaction. 
Since the early XX-th century the observation of different anomalous astrophysical and cosmological effects has led to the introduction of the
Dark Matter hypothesis.
Among the possible explanations is the existence of yet unobserved massive particles, neutral
under electromagnetic interactions.
Depending on 
%the mass of these particles, 
their mass,
they are classified as WIMPs (Weakly Interacting Massive Particles) with masses above 1~GeV/c$^2$~\cite{bib:wimps} or WISPs (Weakly Interacting Slim Particles) with sub-GeV masses~\cite{bib:wisps}. 
%The experimental searches are directed not only towards the hidden sector particles 
These "hidden sector" particles might interact with ordinary matter through a new interaction.
Experimental searches are therefore directed not only to the observation of Dark Matter particles
but also towards the force-carrier of the proposed new interaction.
%There are different hypotheses for such mediator: scalar (e.g. Dark Higgs, with coupling to the Standard model Higgs Boson~\cite{bib:darkhiggs}), pseudo-scalar (e.g. axion portals~\cite{bib:axions}), fermionic (e.g. sterile neutrinos~\cite{bib:sterile}) and vector particles like the Dark Photon $A'$~\cite{bib:darkphoton}.
The models for a vector portal introduce a new U(1) gauge symmetry with its corresponding gauge boson - the Dark Photon $A'$~\cite{bib:darkphoton}. The interaction with the Standard Model fermions 
may arise effectively through the
kinetic mixing term
\begin{equation}
    L_{int} = \frac{\epsilon}{2}F_{QED}^{\mu\nu}F_{\mu\nu}^{Dark},
\end{equation}
where $\epsilon$ is the 
$A-A'$ mixing parameter.
%coupling constant to the electromagnetic interaction.

%PADME (Positron Annihilation into Dark Matter Experiment) is located at Laboratori Nationali di Frascati~\cite{bib:padme}.
The initial setup of PADME (Positron Annihilation into Dark Matter Experiment)~\cite{bib:padme} is designed to probe this hypothesis by searching for the associate production of a Dark Photon together with a visible Standard Model photon in an electron-positron annihilation process:
\begin{equation}
    e^+e^-\rightarrow\gamma A'
\end{equation}
If the visible photon is registered by the experiment's calorimeter, the $A'$ mass can be reconstructed by using the missing mass technique. This requires good knowledge of the main background processes and developing a precise procedure for their rejection by employing several other sub-detectors. This work presents the Dark Photon search strategy for the PADME Run II data as well as a description of the background composition and the steps for its suppression. 

\section{The PADME Run II setup and data taking}
\begin{figure}%[h!]
    \centering 
    \includegraphics[width=0.9\textwidth]{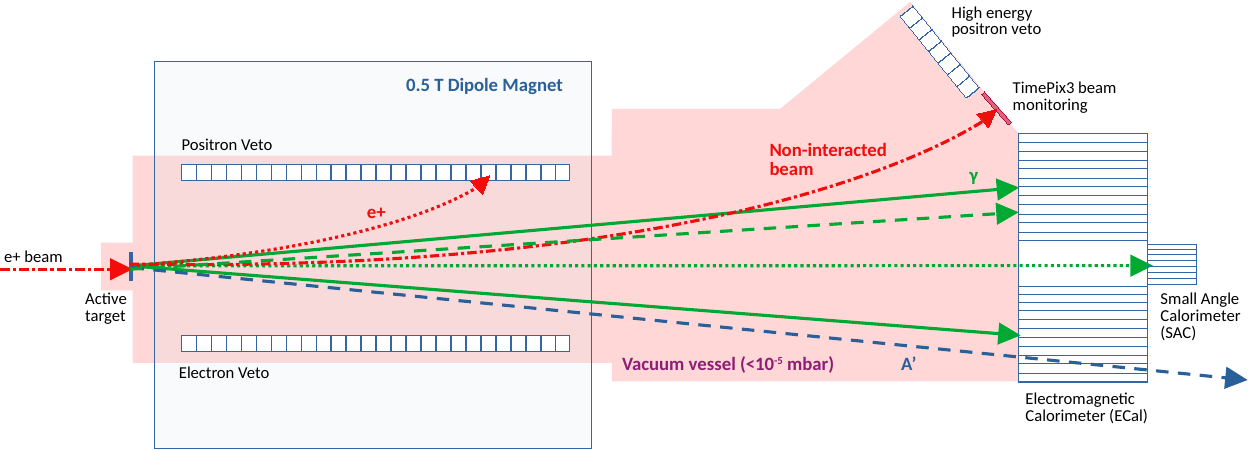}
    \caption[]{PADME Run II setup with several different particles and processes, observed in the experiment. Photons are shown in green, positrons in red and the hypothetical Dark Photon in blue. The continiuous arrows represent two photons, originating from an $e^+e^-\rightarrow\gamma\gamma$ annihilation event; the dashed lines - a photon and a Dark Photon, produced in an $e^+e^-\rightarrow\gamma A'$ event and 
    the
    %with 
    dotted lines 
    %are shown 
    - a positron and a photon, coming from a Bremsstrahlung event.
    %In green: photons, resulting from $\gamma \gamma$ or $\gamma A'$ annihilation events. In red: a positron which emitted a photon through Bremsstrahlung
    }
    \label{fig:padme}
\end{figure}

The PADME experiment uses the pulsed positron beam from the Beam Test Facility (BTF)~\cite{bib:btf} at the DA$\Phi$NE accelerator at Laboratori Nationali di Frascati, capable of reaching energies up to 550 MeV. A schematic view of the experiment with the different subdetectors is shown on Figure~\ref{fig:padme}.
The first detector on the beam line is the active target~\cite{bib:target}
%is a $20~cm\times20~cm\times100~\mu m$ polycrystalline diamond with two sets of 16 graphite electrodes engraved on both its sides, in the x and y direction on the front and back face respectively. That way, apart from 
serving as the medium for the interaction.
%, the target is the first detector in front of the beam, 
It is capable of providing estimation of the beam position and spread, as well as its multiplicity. A dipole magnet provides a 0.5 T magnetic field, bending the non-interacted 
%positrons 
beam
to the 
%beam 
exit window where it is monitored by a TimePix3 silicon pixel detector~\cite{bib:timepix}. 
Three sets of charged particle vetoes~\cite{bib:veto} detect any electrons and positrons which may be part of background events.
They
%segmentation 
provide momentum
measurement with down to $\sim 5$~MeV resolution and $<1$~ns offline time resolution.
%: the positron and electron vetoes inside the magnet
%, two sets of 96 and 90 plastic scintillating bars respectively, 
%and the high energy positron veto next to the beam exit window. 
The calorimetric system is placed at the far end of the apparatus, 3.45~m away from the target.
The Electromagnetic calorimeter (ECal)~\cite{bib:ecal} is made of 616 BGO crystals
and has a $\sim 500$~ps time resolution and $\sim~2.6\%$ energy resolution
at the energy of interest.
The Small Angle Calorimeter (SAC)~\cite{bib:SAC} is placed behind a hole in the middle of the ECal. It provides a $<100$~ps time resolution, important for %dealing with 
registering
the high amount of Bremsstrahlung photons entering this region.

The PADME Run II data taking was realized in the second half of 2020. A primary positron beam with 430 MeV energy was used, with bunch length of 280 ns at a 50 Hz rate and $\sim27\times10^3$ positrons per bunch multiplicity. A total of $\sim5.5\times10^{12}$ positrons were collected during the running period.

\section{Dark Photon search in PADME}

The $A'$ production cross-section $\sigma{(e^+e^-\rightarrow\gamma A')}$  can be obtained by comparing the detected $A'$ events to the estimated number of positrons on target
\begin{equation}
    %N_{A'}=N_{POT}~\sigma{(e^+e^-\rightarrow\gamma A')}~\epsilon_{sig}~N_{e/S}\Rightarrow 
    \sigma{(e^+e^-\rightarrow\gamma A')}=\frac{N_{A'}}{N_{POT}~\epsilon_{sig}~N_{e/S}},
\end{equation}
where $N_{A'}$ is the number of registered A' events, $N_{POT}$ is the number of positrons-on-target, $\epsilon_{sig}$ is the 
acceptance and
%detection efficiency, defined by acceptance, photon registration efficiency and efficiency of the different background cuts,
$N_{e/S} = 0.0106~ b^{-1}$ is the number of electrons per unit of area in the target.
%, $\epsilon$ is the effective coupling to the photon and $\delta$ is a kinematic factor, defining the ratio between $\sigma{(e^+e^-\rightarrow\gamma A')}$ and $\sigma{(e^+e^-\rightarrow\gamma \gamma)}$.
%Assuming kinetic mixing with the electromagnetic interaction, 
Then the effective coupling $\epsilon$ is obtained as
%cross-sections can be used to obtain the effective coupling to the photon $\epsilon$
\begin{equation}
 \epsilon^2 =   \frac{1}{\delta} \frac{\sigma{(e^+e^-\rightarrow\gamma A')}}{\sigma{(e^+e^-\rightarrow\gamma \gamma)}} ,
\end{equation}
where $\delta$ is a kinematics factor depending on $M_{A'}$, with $\delta \to 2$ for $M_{A'}\to 0~\mathrm{MeV}$.

\begin{figure}%[h!]
    \centering      
    \begin{minipage}{0.49\textwidth}
        \centering
        \includegraphics[height=4.5cm]{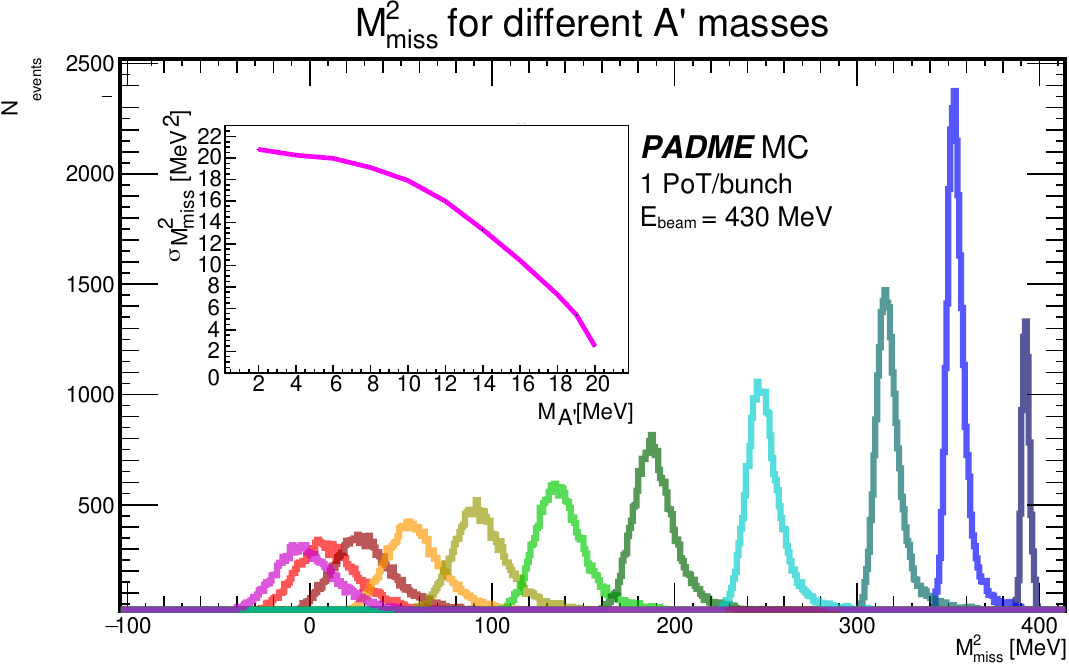}
        \caption{
        %Distribution of the 
        Squared missing mass for photons in simulated $e^+e^-\rightarrow\gamma A'$ events where a single 430 MeV positron hits the target.
        The $\sigma_{M_{miss}^2}$ decreases with $M_{A'}$ approaching 22 MeV.
        %Different $A'$ masses are simulated with the standard deviation $\sigma$ decreasing for higher masses.
        }
        \label{fig:mmiss}
    \end{minipage}\hfill
    \begin{minipage}{0.49\textwidth}
        \centering
        \includegraphics[height=4.38cm]{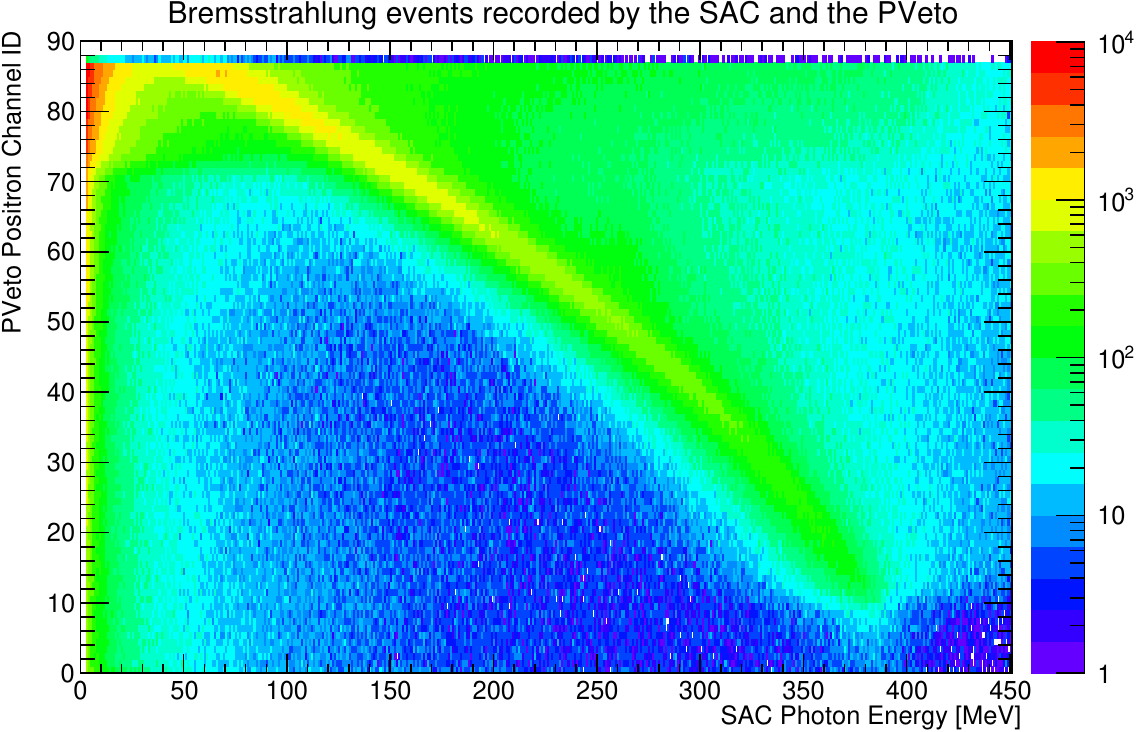}
        \caption{Photon energy in the SAC versus the positron position in the PVeto for in-time $\gamma~e^+$ pairs obtained 
        %, registered in the SAC and PVeto respectively. 
        %The simulation is done 
        for MC simulation
        with $E_{beam}$~=~432~MeV and $2.5\times 10^4$ positrons per bunch.}
        \label{fig:brems}
    \end{minipage}
\end{figure}

%The Dark Photon search technique used by PADME is 
To search for candidate $A'$ events, the missing mass for single photons detected by the experiment's calorimeter is calculated:
\begin{equation}
    M^2_{miss}=(P_{e^+}+P_{e^-}-P_\gamma)^2,
\end{equation}
where $P_{e^+}$, $P_{e^-}$ and $P_\gamma$ are the four-momenta of the beam positron, target electron and registered photon respectively. Figure~\ref{fig:mmiss} shows the distribution of the missing mass squared for different $A'$ masses for events, 
%simulated 
generated
using the PADME Monte Carlo simulation for a single 430 MeV positron hitting the target and producing an $A'$ event.

%The search for an $A'$ signal in PADME Run II data requires reliable background suppression in order to distinguish candidate events. 
%In order to distinguish candidate events, the background events should be precisely filtered.
The main background process is Bremsstrahlung
\begin{equation}
    e^+N\rightarrow e^+N\gamma
\end{equation}
which also results in a single-photon final state. To isolate and suppress these events, photons registered by the calorimeters are matched to positrons, registered by the positron and the high energy positron veto. 
%This is done by matching photon energies to the positions of positrons in the vetoes. 
%A single photon registered by the ECal survives the cut only if there are no matching positrons in 
%a defined isolation time window. 
%time at the expected position in the veto. A Monte Carlo simulation of this distribution for photons registered by the SAC and positrons in the positron veto is shown on Figure~\ref{fig:brems}.
If a photon coincides 
in time ($\Delta t\leq 5~ns$)
with a positron in the veto, 
%at $\Delta t\leq 5~ns$
its energy and the positron position in the veto are checked if they match the Bremsstrahlung distribution (Figure~\ref{fig:brems}), in which case the photon is rejected.

Another background process is two- and three photon annihilation 
\begin{equation}
    e^+e^-\rightarrow \gamma\gamma(\gamma)
\end{equation}
If all the photons are registered in the ECal, these events are rejected by ensuring a time isolation for the selected photon from other photons 
%registered 
in the calorimeters. In 
some
cases only one of the photons is registered,
resulting in a single-photon event.
It can be
%it is 
rejected based on 
its 
%the 
energy and position. 
%distribution 
%for annihilation photons.

%\begin{figure}[h!]
%    \centering
%    \includegraphics[width=0.75\textwidth]{Figures/251215_01.pdf}
%    \caption[]{Brems}
%    \label{fig:brems}
%\end{figure}

%The background suppression procedure is divided in several steps. An initial selection of good photons is performed, based on a minimal energy and geometry acceptance region. A time isolation cut follows and only photons with no other coincidences in a $\Delta t\leq \pm 5~ns$ survive as single photon events. They are additionally filtered to ensure they are not part of incomplete two- and three photon annihilation events. 
%The remaining photons are checked if they 
%belong to the energy-position distribution with positrons, coinciding with them in $\Delta t\leq 5~ns$.
%were produced in Bremsstrahlung events: if a photon coincides with a positron in the veto at $\Delta t\leq 5~ns$, its energy and the positron position in the veto are checked if they match the Bremsstrahlung distribution, in which case the photon is rejected.

\section{Conclusions}
%The second run of the PADME Experiment, realized in 2020, is 
The 2020 run of the PADME Experiment was
dedicated to the search for associate production of a Dark Photon % $A'$ 
%together with a visible photon 
in 
%electron-positron 
$e^+e^-$ annihilation events. The data analysis 
%will utilize 
exploits the missing mass technique, 
%looking for an excess corresponding to $A'$ mass 
probing the region
%between 
$2~\mathrm{MeV} \leq M_{A'}\leq 20~\mathrm{MeV}$. 
%The main obstacle is 
%To suppress the high amount of background events 
%A dedicated procedure was developed 
To cope with the high 
%amount of background event.
background levels, Bremsstrahlung events are suppressed by
%are rejected by matching 
rejecting 
photons matched to close-in-time hits in the positron veto, 
%to positrons, registered by the charged particle detectors 
and 
two- and three photon annihilation events are 
rejected by isolating photons in the calorimeters in time.
Recently, the analysis was boosted forward by the implementation of 
novel machine-learning based methods for pulse and cluster reconstruction~\cite{bib:ML}, 
allowing to cope with the high instantaneous rate.
%is advanced and currently exploits
%If an excess is observed, 
%after the successful application of the procedure
In the case of an excess, 
the number of observed events will allow the estimation of 
%the 
%electromagnetic 
%coupling 
$\epsilon$, 
%of the Dark Photon to the Standard Model photon. I
while in case of no excess, 
%is present, this would put a 
the limit 
%to the observable 
on $\epsilon$ will provide ground for further development of the background reduction procedure 
or for exploring different search techniques.

In 2022 the PADME setup was modified for the search of resonant %X17~\cite{bib:x17} 
production of a hypothetical new particle with 17 MeV mass~\cite{bib:marco}. 
%The magnet was switched off and the non-interacted beam goes straight and through the middle of the ECal. The SAC is removed and the beam is absorbed by a lead glass calorimeter~\cite{bib:katia}. The probing for X17 is done by searching for an excess of $e^+e^-$ couples, registered by the ECal~\cite{bib:marco},~\cite{bib:elisa}.

\section*{Acknowledgements}
%This work is supported by COST Action COSMIC WISPers CA21106, part of COST - European Cooperation in Science and Technology.
The PADME Collaboration acknowledges the support from Istituto Nazionale di Fisica Nucleare, in particular the Accelerator Division and the LINAC and BTF teams of Laboratori Nazionali di Frascati, for providing full support during the data taking.
Sofia University team acknowledges the support by BNSF KP-06-COST/25 from 16.12.2024 based upon
work from COST Action COSMIC WISPers CA21106 and by the European Union - NextGenerationEU, through the National
Recovery and Resilience Plan of the Republic of Bulgaria, project 
SUMMIT BG-RRP-2.004-0008-C01.

\end{document}